\documentclass[conference]{IEEEtran}

\usepackage{amsfonts}
\usepackage{amsmath}
\usepackage{graphicx}
\usepackage{subcaption}
\usepackage{svg}
\usepackage{hyperref,xcolor}

\usepackage{eso-pic}
\newlength{\ieeenoticepullup}
\newcommand{\ieeenoticeadjcol}{\enlargethispage{-\ieeenoticepullup}}
\newcommand{\ieeenotice}{%
  \AddToShipoutPictureFG*{%
    \AtPageLowerLeft{%
      \put(\LenToUnit{\dimexpr1in+\oddsidemargin\relax},\LenToUnit{0.4in}){%
        \makebox[0pt][l]{\parbox[b]{\textwidth}{\fontsize{6.6}{7.6}\selectfont
          \copyright~2025 IEEE. Personal use of this material is permitted. Permission from IEEE must be obtained for all other uses, in any current or future media, including reprinting/republishing this material for advertising or promotional purposes, creating new collective works, for resale or redistribution to servers or lists, or reuse of any copyrighted component of this work in other works.\\
          Accepted version of: Z.~Hu, J.~Tachella, M.~Unser, and J.~Dong, ``Structured Random Model for Fast and Robust Phase Retrieval,'' in \emph{Proc.\ IEEE Int.\ Conf.\ Acoust., Speech Signal Process.\ (ICASSP)}, Hyderabad, India, 2025, pp.~1--5. DOI: \href{https://doi.org/10.1109/ICASSP49660.2025.10889235}{10.1109/ICASSP49660.2025.10889235}.}}}}}}

\newcommand{\signal}{\mathbf{x}}
\newcommand{\measure}{\mathbf{y}}
\newcommand{\sensing}{\mathbf{A}}
\newcommand{\ctrans}{\text{H}}
\newcommand{\overs}{\mathbf{U}}
\newcommand{\unders}{\mathbf{S}}
\newcommand{\fourier}{\mathbf{F}}
\newcommand{\diag}{\mathbf{D}}
\newcommand{\covar}{\mathbf{M}}

\begin{document}
\ieeenotice

\title{Structured Random Model \\ for Fast and Robust Phase Retrieval}

\author{\IEEEauthorblockN{Zhiyuan Hu\IEEEauthorrefmark{1}, Julián Tachella\IEEEauthorrefmark{2}, Michael Unser\IEEEauthorrefmark{1}, Jonathan Dong\IEEEauthorrefmark{1}}
\IEEEauthorblockA{\IEEEauthorrefmark{1}Biomedical Imaging Group, \'Ecole polytechnique f\'ed\'erale de Lausanne, Lausanne, Switzerland }
\IEEEauthorblockA{\IEEEauthorrefmark{2}CNRS, ENS de Lyon, Lyon, France}
}

\maketitle
\ieeenoticeadjcol 

\begin{abstract}

Phase retrieval, a nonlinear problem prevalent in imaging applications, has been extensively studied using random models, some of which with i.i.d.\ sensing matrix components. While these models offer robust reconstruction guarantees, they are computationally expensive and impractical for real-world scenarios. In contrast, Fourier-based models, common in applications such as ptychography and coded diffraction imaging, are computationally more efficient but lack the theoretical guarantees of random models. Here, we introduce structured random models for phase retrieval that combine the efficiency of fast Fourier transforms with the versatility of random diagonal matrices. These models emulate i.i.d.\ random matrices at a fraction of the computational cost. Our approach demonstrates robust reconstructions comparable to fully random models using gradient descent and spectral methods. Furthermore, we establish that a minimum of two structured layers is necessary to achieve these structured-random properties. The proposed method is suitable for optical implementation and offers an efficient and robust alternative for phase retrieval in practical imaging applications.

\end{abstract}

\section{Introduction}
\label{sec:introduction}

Phase retrieval is a long-existing computational problem that finds applications in many areas such as crystallography~\cite{sayre1952some}, astronomy \cite{fienup1987phase},  computer-generated holography \cite{zhang20173d,eybposh2020deepcgh}, optical computing \cite{gupta2019don}, and imaging \cite{zheng2013wide,yeh2015experimental,boniface2020non}. It involves the estimation of a complex-valued vector from magnitude-only measurements. This often leads to a nonlinear optimization problem, which is more challenging than a classical linear regression task. 

Depending on the structure of the sensing matrix, phase retrieval can be categorized into several classes: Fourier, coded illumination, coded detection, and random phase retrieval \cite{dong2023phase}. Among them, the random model is of particular interest. It corresponds to the case in which the elements of the sensing matrix are i.i.d.\ sampled and arises in practical applications such as imaging in complex media \cite{popoff2010measuring,boniface2020non}. Strong theoretical guarantees have been derived for this random setting: we know when the problem is theoretically solvable~\cite{bandeira2014saving,mondelli2019fundamental,maillard2020phase} and which algorithm to use \cite{maillard2020phase, luo2019optimal}. This is in sharp contrast with the other settings of phase retrieval in which many theoretical questions remain unsolved. For these reasons, the broader deployment of the random model has the potential to unlock robust phase retrieval in practice. However, building a dedicated system for high-performance phase imaging is unrealistic as it would involve a prohibitively large sensing matrix to characterize and use. Interestingly, acceleration techniques to replace random-matrix multiplications have been proposed for feedforward \cite{yu2016orthogonal} and recurrent \cite{dong2020reservoir,d2025comparison} neural networks. These structured random matrices emulate the randomness of i.i.d.\ random matrices using Hadamard transforms and random diagonal matrices. 

\ieeenoticeadjcol 
In this work, we introduce the concept of structured random models for phase retrieval and show that they provide a computationally efficient alternative for fast and robust phase reconstruction. With structured layers combining Fourier transforms and random diagonal matrices, this structured random model reduces the computation time of the forward and backward pass from $\mathcal{O}(n^2)$ to $\mathcal{O}(n\log n)$, as well as memory complexity from $\mathcal{O}(n^2)$ to $\mathcal{O}(n)$. We show that the structured random model enables the reconstruction of signals at a quality that is similar to that of the original i.i.d.\ random model. We also investigate the optimal number of structured and diagonal matrices, showing that a cascade of two is optimal. Our structured random model can be efficiently parallelized on GPU and could potentially be implemented in optical microscopy \cite{goodman2005introduction, moczulski2016acdc}. 

\vspace{1em}

\section{Background}
\label{sec:background}

\subsection{Random Phase Retrieval}
In phase retrieval, one seeks to recover the $n$-dimensional complex signal $\signal \in \mathbb{C}^n$ by measuring the $m$-dimensional real signal $\measure \in \mathbb{R}^m$ through a nonlinear model,
\begin{equation}
    \measure = | \sensing \signal |^2,
\end{equation}
where $\sensing \in \mathbb{C}^{m \times n}$ is a known matrix and $|\cdot|$ represents the component-wise modulus operation. For random phase retrieval, the elements of $\sensing$ are i.i.d.\ sampled from a given distribution. An important hyperparameter is the oversampling ratio $\alpha = m / n$: larger oversampling ratios provide more information about the signal which makes the reconstruction problem simpler.

Among the rich literature of theoretical investigations on complex-valued i.i.d. random phase retrieval, we can mention a few notable asymptotic results. First, the mapping between the signal $\signal$ and the measurement $\measure$ is injective for oversampling ratio $\alpha > 4$~\cite{bandeira2014saving}. Moreover, the weak-recovery threshold above which one can start to reconstruct better than random guess is $\alpha_{\text{WR}} = 1$ \cite{mondelli2019fundamental}, and the full recovery threshold is $\alpha_{\text{FR}} = 2$~\cite{maillard2020phase}. The nonlinear inverse problem has been well characterized in the random setting with efficient algorithms to solve it with theoretical guarantees. 

Randomized models have been introduced as a way to improve reconstruction performance. Two special schemes are: (i) coded diffraction imaging, which employs random masks followed by a Fourier transform \cite{candes2015phase_wf}; and (ii) random-probe ptychography \cite{pelz2017low}, which uses a shifted random probe. However, the robust reconstruction results of random phase retrieval do not directly apply in both cases. 

\subsection{Reconstruction Algorithms}
\label{sec:reconstrucion_algoritms}

\subsubsection{Gradient Descent}
from an optimization perspective, gradient descent is a classic yet powerful method to solve the reconstruction task. Using the vanilla $\ell_2$ loss $\mathcal{L}(\hat{\signal}) = \| |\sensing\hat{\signal}|^2 - \measure \|^2$, the gradient of the loss is given by
\begin{equation}
\label{eq:l2_gradient}
    \mbox{{\boldmath{$\nabla$}}} \mathcal{L}_{\hat{\signal}} = - 2 \sensing^\ctrans\mathbf{diag}(\hat{\signal})(\measure-|\sensing \hat{\signal}|^2),
\end{equation}
where $(\cdot)^\ctrans$ denotes conjugate transpose and $\mathbf{diag}(\cdot)$ denotes the diagonal matrix formed using the given vector. We observe that \eqref{eq:l2_gradient} involves both the forward matrix $\sensing$ and its adjoint $\sensing^\ctrans$. Other choices of losses are also possible and may be deduced from noise statistics \cite{yeh2015experimental}.

Initialization is key to the performance of the gradient-descent algorithm~\cite{candes2015phase_wf}. Due to the nonlinearity of the forward model, the loss landscape presents many local minima, which explains why random initialization often leads to unstable performance.

\subsubsection{Spectral Methods}
Spectral methods have emerged as an appealing option with low computational cost and competitive performance. The idea is to estimate the reconstruction signal $\hat{\signal}$ as the leading eigenvector of the weighted covariance matrix
\begin{equation}
\label{eq:covariance_matrix}
    \covar(\mathcal{T}) = \frac{1}{m} \sum_{i=1}^m \mathcal{T}(y_i) \mathbf{a}_i \mathbf{a}_i^\ctrans,
\end{equation}
where $y_i$ is the $i$th element of the measurement $\measure$, $\mathbf{a}_i^\ctrans$ is the $i$th row of the sensing matrix $\sensing$, and $\mathcal{T}$ is an increasing preprocessing function. The preprocessing function $\mathcal{T}(y) = \text{max}\{1 - 1/y, \lambda\}$ has been proven to be optimal for noiseless i.i.d.\ random phase retrieval, where $\lambda < 0$ is a lower bound threshold \cite{luo2019optimal,ma2021spectral,maillard2021construction}.

Power iterations are typically used to obtain this leading eigenvector, initialized with a random vector $\hat{\signal} \in \mathbb{C}^n$ and repeatedly updated through the matrix multiplication
\begin{equation}
    \hat{\signal}_{\text{new}} = \covar \hat{\signal} - 2\lambda \hat{\signal},
\end{equation}
where $\covar$ is the covariance matrix defined in \eqref{eq:covariance_matrix}.

While spectral methods are well suited to the random setting, they provide only modest improvements for nonrandom models in general~\cite{valzania2021accelerating}.

\subsubsection{Other Reconstruction Methods}
Apart from gradient descent and spectral methods, a few other notable resolution schemes are projection algorithms~\cite{fienup1982phase}, convex relaxations~\cite{candes2013phaselift}, and Bayesian approximate message passing~\cite{baker2020tramp,celentano2020estimation}. The latter is known to provide the most accurate results but is also the most complex and difficult to implement efficiently for large-scale imaging problems.

\vspace{1em}

\section{Structured Random Phase Retrieval}
\label{sec:pseudonrandom_phase_retrieval}

We define structured random sensing matrices with cascaded structured layers $\fourier \diag$ as
\begin{equation}
\label{eq:peudorandom}
\sensing = \begin{cases}
\unders \prod_{i=1}^N (\fourier \diag_i ), \quad \alpha \leq 1,\\ 
\prod_{i=1}^N (\fourier \diag_i ) \overs, \quad \alpha > 1,
\end{cases}
\end{equation}
where $\fourier$ is the FFT matrix and $\diag_i$ are complex random diagonal matrices with elements of unit magnitude and random phase sampled from $\text{Unif}_{[0,2\pi)}$.

The matrices $\unders, \overs \in \mathbb{C}^{m\times n}$ represent the subsampling and oversampling operations, respectively. We achieve them by two different physically relevant approaches: for oversampling, we zero-pad the images to enlarge the dimension of the original signals and then multiply by $\fourier \diag_i$ to acquire the oversampled measurements; for undersampling, we first apply $\fourier \diag_i$ on the original image, and then trim the edges of the signals to decrease the output dimension. 

The intuition for this process is as follows: A multiplication by an i.i.d.\ random matrix $\sensing$ performs a random all-to-all mixing of the components of $\signal$. The structured random model emulates this behavior by two operations: The FFT operation mixes all components of the signal efficiently, while the random diagonal matrix adds random phase factors. Combined, these two operations mimic a random i.i.d.\ matrix multiplication.

The computational complexity for the forward pass of the structured random model is $\mathcal{O}(n\log n)$ as driven by the FFT, in contrast to $\mathcal{O}(n^2)$ for a fully unstructured model.
\begin{figure}[t]
    \centering
    \includesvg[width=1.0\linewidth]{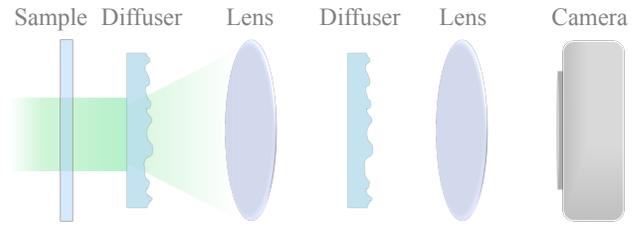}
    \caption{Optical system implementing the structured random model.}
    \label{fig:optics}
\end{figure}

The proposed sensing matrix can be implemented physically using optical components. Based on the Fourier-optics formalism~\cite{goodman2005introduction}, each Fourier transform can be implemented as a lens, while each diagonal matrix can be realized as a random diffuser---a rough glass surface that imprints a random phase pattern onto the optical wavefront. Thus, the structured sensing matrix with 2 diagonal matrices can be physically realized through a sequence of optical elements as illustrated in Fig.~\ref{fig:optics}: first a diffuser, followed by a lens, then a second diffuser, and a final lens. Such optical systems bear similarities to other diffuser-based imaging systems such as DiffuserCam~\cite{antipa2017diffusercam}. However, it is worth noting that, while these systems have been applied in various imaging contexts, they have not been applied to phase imaging so far.

\section{Results}
\label{sec:results}
In this section, we present numerical results to demonstrate the feasibility of the structured random model and to highlight its advantages. First, we run different reconstruction algorithms on i.i.d.\ random and structured random models to compare their performance. Second, we compare the performance of structured random models with different numbers of layers. Last, we benchmark the forward time of the two models on both CPU and GPU.

\urlstyle{tt}

We implement i.i.d.\ random and structured random phase retrieval in Python. Since we focus on the image-processing domain, the Fourier transform operation is performed using a 2D FFT. For spectral methods, we use the optimal preprocessing function as introduced in Section \ref{sec:background}. The models and corresponding reconstruction algorithms are implemented in the open-source image-processing library DeepInverse \cite{Tachella_DeepInverse_A_deep_2023} and the source code to generate the figures is publicly available\footnote{\small \url{https://github.com/zhiyhucode/structured\_random\_phase\_retrieval}}.

\subsection{Reconstruction Performance}

In phase retrieval, it is standard to use the cosine similarity between the original signal and the reconstruction. This corresponds to the following precision metric
\begin{equation}
    \text{cosine similarity} = \frac{|\langle \hat{\signal},\signal^* \rangle|}{\|\hat{\signal}\| \cdot \|\signal^*\|},
\end{equation}
where $\signal^*$ is the original signal, $\hat{\signal}$ is the reconstruction, and $\langle \mathbf{a},\mathbf{b} \rangle = \mathbf{a}^\ctrans \mathbf{b}$ is the complex inner product. Cosine similarity yields a scalar in $[0,1]$, with 0 for an uncorrelated estimate and 1 for a perfect recovery.

To generate the benchmark signal, we choose the standard test image Shepp-Logan phantom \cite{shepp1974fourier} and map its grayscale pixel values in $[0,1)$ linearly to $[-\pi/2,\pi/2)$ as the phase of complex-valued signals with unit magnitude. Three reconstruction methods are used for both models: (1) gradient descent with random initialization (GD); (2) spectral methods only (SM); and (3) gradient descent with spectral initialization (GD + SM). We always use 10000 gradient-descent iterations (with early stopping) and 5000 power iterations for spectral methods (with early stopping). We rely on these large numbers of iteration to ensure the algorithm convergence. Besides, we perform 100 repetitions to obtain sufficient statistics and plot the error bars with 10\%/90\% quantiles.
\begin{figure}[t]
\begin{minipage}[b]{1.0\linewidth}
  \centering
  \centerline{\includesvg[width=1.1\linewidth]{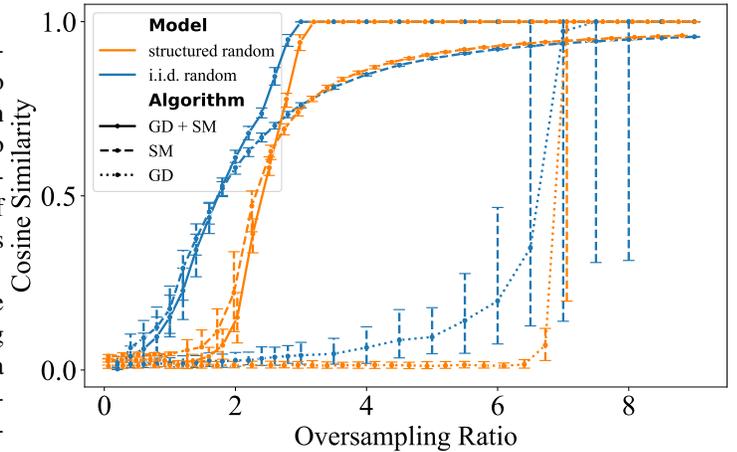}}
  \subcaption{Reconstruction accuracy for structured random models and i.i.d.\ random models.}
  \label{fig:accuracy}
\end{minipage}
\begin{minipage}[b]{1.0\linewidth}
  \centering
  \centerline{\includesvg[width=1.2\linewidth]{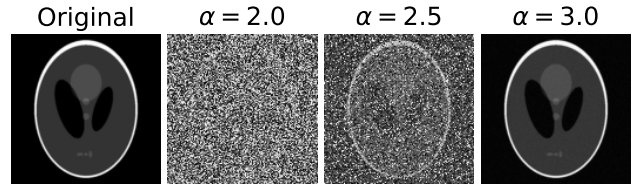}}
  \subcaption{Reconstruction from structured random models for different oversampling ratios.}
\end{minipage}
\caption{Reconstruction performance of structured random models. Structured random models achieve comparable performance with i.i.d. random models and a perfect recovery at an oversampling ratio of 3.}
\end{figure}
In Fig.\ \ref{fig:accuracy}, we display the cosine similarities for the three aforementioned algorithms as a function of the oversampling ratio. It can first be seen that, for the i.i.d.\ random model, gradient descent with random initialization is subject to high variance and low accuracy, only achieving a good recovery for very high oversampling ratio (around 8). In contrast, spectral methods provide a much more stable and accurate reconstruction. Followed by gradient descent, full recovery is achieved around an oversampling ratio of 3.

For the structured random model with 2 layers, the curves are similar to those of the i.i.d.\ random model while being typically steeper. For gradient descent combined spectral methods, the curve only starts rising when the oversampling ratio gets above 1.5, but still quickly improves and matches the performance of the i.i.d.\ random model at an oversampling ratio of 3. 

The performance improvement of the structured random model also aligns with theoretical analysis. As shown in~\cite{maillard2020phase}, the weak and strong recovery thresholds for unitary sensing matrices are 2 in the infinite scenario.  With the structured random sensing matrix being unitary, we therefore see a substantial increase above oversampling 2. Another piece of evidence is that the variance greatly decreases above an oversampling ratio of 2, which indicates that the reconstruction becomes more robust.

\vspace{.5em}

\subsection{Number of Layers}
To understand the role of the structured layer $\fourier \diag$ in the structured random model, we run the spectral methods for 5000 iterations alone and with a following gradient descent for structured random models with 1 to 3 layers from oversampling ratios 2 to 5. We also run the same algorithms on models with random unitary sensing matrices.
\begin{figure}[t]
  \centering
  \centerline{\includesvg[width=1.05\linewidth]{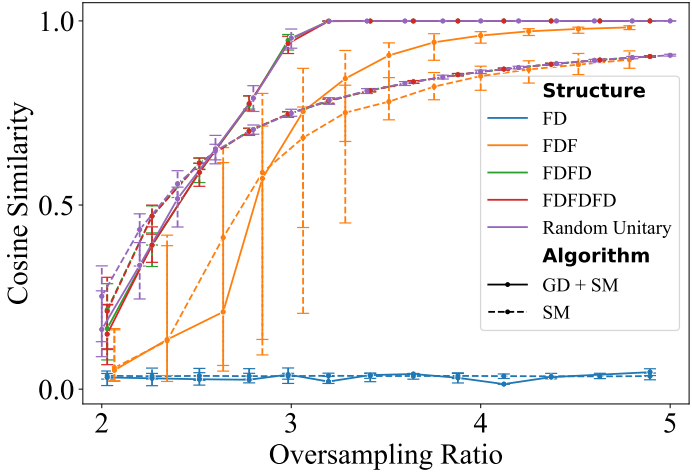}}
\caption{Performance of structured random models with several architectures and random unitary models. Two structured layers are sufficient to achieve the best results.}
\label{fig:layers}
\end{figure}
We show in Fig.\ \ref{fig:layers} the results of these experiments. We see that the structured random model demonstrates a substantial performance gain by employing 2 layers instead of 1 but has a saturated performance with more than 2 layers, under which perfect recovery is achieved around an oversampling ratio of 3.2. Besides, by adding an extra FFT to the 1-layer model, we obtain results that are intermediate between an uncorrelated guess and the optimal performance.

Another interesting phenomenon is that the performance of structured random models with more than 2 layers is very similar to the one of random unitary matrices. This can be related to the similarity between the spectral distributions of structured random and random unitary sensing matrices, since it is shown in \cite{mondelli2019fundamental, maillard2021construction} that matrices with similar spectral distributions will yield similar results with the same algorithm.

These phenomena indicate that it is efficient to use 2 layers to achieve the desired performance. This result provides important insights for future experimental implementation, as minimizing the number of layers would reduce experimental complexity. 

\vspace{.5em}

\subsection{Computational Complexity}
\begin{figure}[t]
  \centering
  \centerline{\includesvg[width=1.0\linewidth]{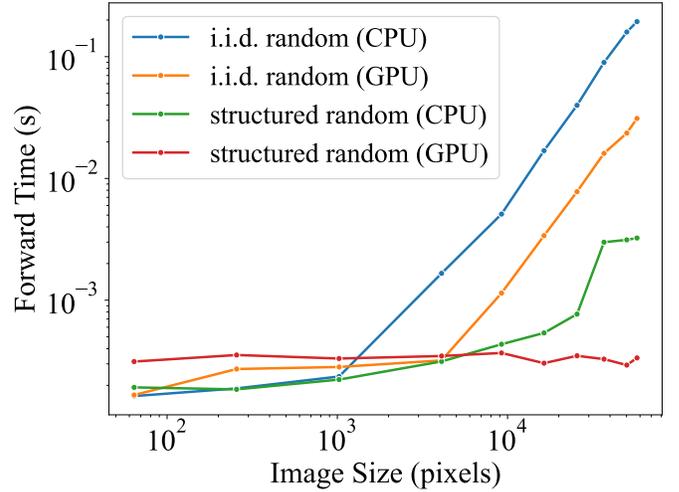}}
\caption{Time benchmark of structured random and i.i.d.\ random models on CPU and GPU. Structured random models are substantially faster than i.i.d. random models on both CPU and GPU.}
\label{fig:time}
\end{figure}

We also benchmark the forward time of the i.i.d.\ and structured random models on both CPU and GPU with a fixed oversampling ratio of 1, as shown in Fig.\ \ref{fig:time}. Thanks to the parallelization of GPUs, the structured random model can achieve a constant forward time despite the increase in image size smaller than $10^6$ pixels, even better than the asymptotic $O(n\log n)$. For image sizes larger than $10^6$ pixels, the forward time of structured random models on GPU also starts to increase. In comparison, the processing time for the i.i.d.\ random model grows substantially when the image size increases. Moreover, it is also time-consuming to sample the i.i.d.\ weights when the matrix becomes large.

For results on CPU, we see a quadratic increase for the forward time of the random model. This is in good agreement with the theoretical complexity, while the structured random model is much more efficient but still becomes considerably large with large image sizes. Besides, the simulation for the i.i.d. random model is limited to the largest image size of ($240 \times 240$) due to the memory limit of the GPU to store the full matrix.

\vspace{1em}

\section{Conclusion}
\label{sec:conclusion}

We propose a novel imaging model for phase retrieval. It efficiently incorporates randomness by cascading FFTs and random diagonal matrices. Our approach leads to reconstruction performance that is similar to that of an unstructured random model, allowing for a full recovery with an oversampling ratio of 3. Moreover, our proposed approach requires considerably less processing time than random i.i.d.\ or unitary matrices, more specifically, a constant forward time on GPU for moderate image sizes.

Interesting directions for future work to understand better these structured random models may involve: the mechanism behind the performance leap from 1 layer to 2 layers; the set of structures that yield structured randomness, and the influence of using different sampling distributions. For practical applications, hardware experiments are possible using lenses to implement FFTs and diffusers for the random elementwise multiplications. 

\vspace{1em}

\section{Funding}

Zhiyuan Hu and Jonathan Dong acknowledge funding from the Swiss National Science Foundation (Grant PZ00P2\_216211). Julián Tachella is supported by the ANR grant UNLIP (ANR-23-CE23-0013).

\vfill\pagebreak

\bibliography{refs}

\bibliographystyle{IEEEtrans}

\end{document}